\documentclass[12pt,a4paper]{article}
\usepackage{amsfonts,amssymb,amsmath,amsthm,cite}
\usepackage{graphicx}
\usepackage{slashed}

\evensidemargin=\oddsidemargin
\newcommand{\Tr}{\mathrm{Tr}\, }
\newcommand{\ii}{\mathrm{i}}

\begin{document}
\title{{\bf Fractional Fermion Number \\ and \\ Hall Conductivity of Domain Walls}}

\author{ J. Mateos Guilarte$^{1}$\footnote{guilarte@usal.es} and D. Vassilevich$^{2,3}$\footnote{dvassil@gmail.com}
\\
\small $^{1}$  Departamento de Fisica Fundamental, University of Salamanca, Spain \\
\small $^{2}$ CMCC, Universidade Federal do ABC, Santo Andre, Sao Paulo, Brazil\\
\small $^{3}$ Physics Department, Tomsk State University, Tomsk, Russia
}

\date{\today}

\maketitle

\begin{abstract}
In this letter the fractional fermion number of thick domain walls is computed. The analysis is achieved by developing the heat kernel expansion of the spectral eta functon of the Dirac Hamiltonian governing the fermionic fluctuations around the domain wall. A formula is derived showing that a non null fermion number is always accompanied by a Hall conductivity induced on the wall. In the limit of thin and impenetrable walls the chiral bag boundary conditions arise, and the Hall conductivity is computed for this case as well.
\end{abstract}

\section{Introduction}\label{sec:Intro}
The possibility of a fractional fermion number of solitons was first noted by Jackiw and Rebbi in 1976 \cite{Jackiw:1975fn}. The main results and ideas of the first decade of the development in this area were reported in \cite{Niemi:1984vz}. From the early days, condensed matter physics was among the main applications \cite{Jackiw:1981wc}. No wonder that the advances in Dirac materials caused a renewed interest to the charge fractionization, see e.g. \cite{Chamon:2007hx,Wang:2017xhg,Yang:2019}. New applications required new technical tools. In particular, an efficient method for calculation of the fractional charge based on a resummation of the heat kernel expansion was suggested recently in \cite{Alonso-Izquierdo:2019tms}.

In the present work we consider domain walls in $(3+1)$ dimensions that separate regions in the space characterized by asymptotically constant values of a scalar and an axial scalar fields. There is magnetic flux through the wall. As we shall see, the resummation method of \cite{Alonso-Izquierdo:2019tms} works straightforwardly  and relatively easy in this case and produces a nice formula for the fractional charge in terms of the total magnetic flux and of the chiral angle of scalar fields in the asymptotic regions.

The fermion number fractionization is connected with quantum anomalies and with the parity anomaly \cite{Niemi:1983rq,Redlich:1983dv} in particular. It was believed for a long time that there cannot be a parity anomaly in $(3+1)$ dimensions. However, recent calculations demonstrated \cite{Kurkov:2017cdz,Kurkov:2018pjw} that on manifolds with boundaries the parity anomaly exists also in four dimensions and leads to an induced Chern-Simons interaction on the boundary. A computation of the Chern-Simons term on a domain wall was done in \cite{Mulligan:2013he}. One of the  main physical motivations for these computations was to clarify if there is a Hall conductivity on the boundary of a Dirac material. Our results on the fermion number allow to fix the induced Chern-Simons action for domain walls. Since chiral bag boundary conditions may be obtained as a strong coupling limit of domain walls, we are able to compute the Chern-Simons term on the boundaries as well. 

\section{Fermion number}\label{sec:fer}
In this article we consider quantized Dirac fermions in $(3+1)$ dimensions interacting with background scalar and axial scalar fields, $\varphi_1$ and $\varphi_2$, and an electromagnetic potential $A_\mu$. The Lagrangian of this system reads
\begin{eqnarray}
&&\mathcal{L}=\bar\psi \slashed{D}\psi,\qquad
\slashed{D}=\ii \gamma^\mu D_\mu -\varphi_1 -\ii \gamma^5 \varphi_2, \qquad D_\mu=\partial_\mu -\ii e A_\mu, \label{Dirop}\\ &&\gamma^j=\beta\alpha^j \, , j=1,2,3 \, \, \, , \, \, \, \gamma^0=\beta \, \, \, , \, \, \gamma^5=\ii\gamma^0\gamma^1\gamma^2\gamma^3 .\nonumber 
\end{eqnarray}
Here $e$ denotes the elementary charge, and $\alpha^j$, $\beta$ are the standard Dirac matrices. Let $\epsilon^{\mu\nu\rho\sigma}$ be the Levi-Civita tensor. Then,
\begin{equation*}
\mathrm{tr}\, \bigl( \gamma^5\gamma^\mu\gamma^\nu\gamma^\rho\gamma^\sigma\bigr)=-4\ii\epsilon^{\mu\nu\rho\sigma},\qquad \epsilon^{0123}=1.
\end{equation*}

If the background is static, the Dirac Hamiltonian reads
\begin{equation}
H=-\ii \alpha^j D_j +\beta \varphi_1 +\ii \beta \gamma^5 \varphi_2 \,.\label{DirH}
\end{equation}

Let $\lambda$ denote the eigenvalues of $H$. Then the spectral $\eta$ function is defined by a sum over the spectrum
\begin{equation}
\eta (s,H)=\sum_{\lambda >0} \lambda^{-s} - \sum_{\lambda <0} (-\lambda)^{-s}, \label{zeta0H}
\end{equation}
which is convergent if $\Re s$ is sufficiently large, and may be extended to the whole complex plane as a meromorphic function. The value of the $\eta$ function at $s=0$ defines the fermion number in the vacuum \cite{Niemi:1984vz}, 
\begin{equation}
N=-\tfrac{1}{2} \eta(0,H)
\end{equation}
To evaluate the spectral asymmetry of this Hamiltonian and thus the vacuum fermion number we use the methods elaborated in \cite{Alonso-Izquierdo:2019tms}. Let us take a smooth function $\rho$ of compact support and define a localized $\eta$ function,
\begin{equation}
\eta (s,H;\rho)= \Tr \left( \rho \cdot (H^2)^{-s/2} H/|H| \right) 
= \frac 1{\Gamma \left( \frac{s+1}2 \right)} \int_0^\infty dt\, t^{ \frac{s-1}2}
\Tr \left( \rho H e^{-tH^2} \right) .\label{etarho}
\end{equation}
With the help of the localized $\eta$ function we can define a current $j^0$, that gives the global fermion number $N$ after the integration
\begin{equation}
\eta(0,H;\rho)=-\frac 12 \int d^3x j^0(x) \rho(x),\qquad \int d^3 x j^0(x)=N \label{etaj0}
\end{equation}
Note that locally $j^0(x)$ does not need to coincide with the charge density.
The function $\eta(s,H;\rho)$ may be expressed as
\begin{equation}
\eta (s,H;\rho)=-\frac 1{2\Gamma \left( \frac{s+1}2 \right)} 
\int_0^\infty dt\, t^{ \frac{s-3}2} \frac {\mathrm{d}}{\mathrm{d}\varepsilon} \vert_{\varepsilon=0} \, \Tr \left(  e^{-tH_{\rho}^2} \right) \label{eta3}
\end{equation}
where
\begin{equation}
H_\rho =H+\varepsilon \rho \label{Hrho}
\end{equation}

Let $L$ be a Laplace type operator. It can be represented in the canonical form
\begin{equation}
L=-(\nabla^2+E)\label{Lap}
\end{equation}
with some matrix valued potential $E$ and a covariant derivative $\nabla=\partial +\omega$. We shall be interested in
\begin{equation}
L(\rho,M^2)=H_\rho^2-M^2. \label{LrM}
\end{equation} 
Here $M$ is an auxiliary mass parameter that is needed to organize a derivative expansion of the $\eta$ function (see \cite{Alonso-Izquierdo:2019tms} for details). For this operator,
\begin{eqnarray}
&&E=-\tfrac {\ii e}4 F_{jk}[\gamma^j,\gamma^k] -\ii \gamma^j\partial_j\varphi_1 +
\gamma^j\gamma^5\partial_j\varphi_2 +(M^2 - \varphi_1^2-\varphi_2^2)-2\beta\varepsilon\rho (\varphi_1 +\ii \gamma^5\varphi_2) \nonumber\\
&&\omega_j =-\ii e A_j +\ii \alpha^j \varepsilon \rho \label{EoO}\\
&& \Omega_{jk}\equiv [\nabla_j,\nabla_k]=-\ii eF_{jk} +\alpha_k \varepsilon \partial_j\rho - \alpha_j \varepsilon \partial_k \rho \nonumber
\end{eqnarray}
Any Laplace type operator admits an asymptotic expansion of the heat trace (the heat kernel expansion)
\begin{equation}
\Tr \left(Qe^{-tL}\right)\simeq \sum_{k=0}^\infty t^{\frac{k-3}2}a_k(L,Q), \qquad t\to +0 \label{hke}
\end{equation}
with any smooth matrix valued function of rapid decay. On manifolds without boundaries all coefficients with even $k$ vanish. Here we shall need just a couple of basic facts \cite{Vassilevich:2003xt} about the heat kernel coefficients $a_k$: (i) they all are integrals of traces of local polynomials constructed from $E$, $\Omega$ and their covariant derivatives $\nabla$ with vector indices contracted in pairs, and (ii) the terms depending on $E$ only have the following simple form:
\begin{equation}
a_{2l}(L,Q)\sim \frac 1{(4\pi)^{\frac 32}l!} \int d^3x \mathrm{tr}\, \left( QE^l \right).\label{a2l}
\end{equation}
In $a_2$, there are no other contributions, so that (\ref{a2l}) is exact for $a_2$.
 
Let us return to our specific problem. Consider arbitrary small localized variations of the background fields, $\delta A$ and $\delta\varphi_{1,2}$. Then, $\delta H=- \alpha^j\delta A_j +\beta\delta\varphi_1+\ii\beta\gamma^5\delta\varphi_2$. The resulting variation of $\eta(0)$ reads \cite{Gilkey,AlvarezGaume:1984nf}
\begin{equation}
\delta \eta(0,H)= -\frac 2{\sqrt{\pi}} a_{2}(H^2,\delta H)\,,\label{varEta}
\end{equation}
By using (\ref{EoO}) and (\ref{a2l}) one easily computes $\delta\eta(0,H)=0$. Therefore, $\eta(0,H)$ is a topological (or homotopy) invariant.

The fermion number $N$ will be evaluated through the large mass/small derivatives expansion with respect to the parameter $M^2$.
\begin{equation}
\eta (0,H;\rho)= -\frac 1{2\sqrt{\pi}} \sum_k  \Gamma\left( \frac{k}2-2 \right)
|M|^{4-k} \frac {\mathrm{d}}{\mathrm{d}\varepsilon} \vert_{\varepsilon=0} \, a_k\bigl(L(\rho,M^2)\bigr) .\label{largeM}
\end{equation}

Let us consider the field configurations for that the only nonvanishing component of $F_{ij}$ is $F_{12}$ that does not depend on $x^3$. We assume that $\varphi_1$ and $\varphi_2$ do not depend on $x^1$ and $x^2$ and go exponentially fast to their asymptotic values at $x^3\to \pm\infty$. We shall restrict ourselves to the first order of $F_{12}$. As in \cite{Alonso-Izquierdo:2019tms}, we need to keep in the heat kernel coefficients only the terms that contain at most one derivative w.r.t. $x^3$ and no derivatives w.r.t. other coordinates. Besides, such terms have to be linear in $\varepsilon\rho$. It is easy to see, that in $a_{2p}$ the right number of derivatives may be obtained in the combinations $(\nabla E)^2 E^{p-3}$, $\Omega_{ij}\Omega^{ij} E^{p-2}$ and $E^p$ (with a possibility to reorder the multiples). A more attentive analysis of the traces over $\gamma$-matrices shows that in fact only the $E^{k/2}$ invariants contribute. The corresponding heat kernel coefficients read:
\begin{eqnarray}
&&a_{2(l+3)}\sim \frac 1{(4\pi)^{\frac 32}l!} \int d^3x\, \mathrm{tr}\left\{ (M^2-\varphi_1^2-\varphi_2^2)^l (\ii e F_{12}\gamma^1\gamma^2)\right.\nonumber\\
&&\qquad\qquad\qquad\qquad \left.\times \left[ (-\ii\gamma^3\partial_3\varphi_1) (-2\ii\beta \varepsilon\rho \gamma^5 \varphi_2) + (\gamma^3\gamma^5\partial_3\varphi_2)(-2\beta\varepsilon\rho\varphi_1)\right]\right\}\nonumber\\
&&\qquad\quad =\frac {8e\varepsilon}{(4\pi)^{\frac 32}l!} \int d^3x\, (M^2-\varphi_1^2-\varphi_2^2)^l F_{12} (\varphi_2\partial_3\varphi_1 -\varphi_1\partial_3\varphi_2)\rho \label{a}
\end{eqnarray}
We substitute (\ref{a}) in (\ref{largeM}) and sum over $l$ to obtain
\begin{equation}
N=-\frac {e}{4\pi^2} \left.\left( \mathrm{arctg} (\varphi_2/\varphi_1) \right)\right\vert_{x^3=-\infty}^{x^3=+\infty} \int d^2x F_{12} \,.\label{Nfin}
\end{equation}
The sum is convergent as long as $\varphi_1^2+\varphi_2^2\neq 0$.

This result admits an elegant interpretation in terms of the chiral angle. Let
\begin{equation}
\varphi_1=\varphi \cos \theta,\quad \varphi_2=\varphi \sin \theta, \quad
\varphi=\sqrt{\varphi_1^2+\varphi_2^2},\quad \theta =\mathrm{arctg}(\varphi_2/\varphi_1).\label{theta}
\end{equation}
Then, $N$ is proportional to $\theta^+ -\theta^-$ with $\theta^\pm \equiv \lim_{x^3\to\pm\infty} \theta(x^3)$. The Lagrangian (\ref{Dirop}) is invariant under the global chiral rotations
\begin{equation}
\theta (x) \to \theta(x)+\delta\theta,\qquad \psi \to \exp\left( -\tfrac {\ii}2 \delta\theta\gamma^5 \right)\psi .\label{chi}
\end{equation}
This symmetry is broken in quantum theory due to the anomaly. However, the fermion number $N$ remains invariant under the global chiral rotations.

The fermion number (\ref{Nfin}) is a product of two terms. One is the magnetic flux through the $(x^1,x^2)$ plane reminding us of the index of a two-dimensional Dirac operator. The other - is the fermion number in the Goldstone-Wilczek model \cite{Goldstone:1981kk} in $(1+1)$ dimensions. For compact manifolds, the factorization properties may be demonstrated on general grounds \cite{Gilkey}. In the non-compact case, the situation is more complicated, though a similar factorization structure of the fermion number was demonstrated for the magnetic monopole \cite{Paranjape:1983dy,Niemi:1984vz}.

If the fields $\varphi_1$ and $\varphi_2$ remind the profiles of solitons (i.e., if they go fast to their asymptotic values at $x^3\to \pm \infty$), both $\theta^+$ and $\theta^-$ are well defined. A problem may apparently appear if $|\varphi|=0$ for $x^3>X$ with some $X$ (or in a similar situation in a vicinity of $-\infty$). Then, for $x^3>X$ also $\partial_3\varphi_1=\partial_3\varphi_2=0$ so that by Eq.\ (\ref{a}) this region does not contribute to the fermion number. Thus,in (\ref{Nfin}) the upper limit has to be taken $\theta^+=\lim_{x^3\to Y}\theta(x^3)$ where $Y$ is the upper bound of $\{ x^3\}$ such that $|\varphi(x^3)|\neq 0$.

\section{Induced Chern-Simons term on an interface}\label{sec:ind}
In this section, we evaluate the Chern-Simons action induced at an interface by quantum effects. We have to consider non-static backgrounds and thus to work effective actions in 4D. Previously this problem was considered in \cite{Mulligan:2013he}. 

Let us consider a domain wall background defined by the fields $\varphi_{1,2}$ that depend on $x^3$ only. Suppose that these fields change rapidly for $x^3$ near zero and go exponentially fast to the asymptotic values away form $x^3=0$. At the beginning, we do not impose any restrictions on the external electromagnetic field $A_\mu$. The one-loop effective action for spinors restricted to the second order in $A_\mu$ contains two contributions, the parity even and parity odd ones. Roughly speaking, the parity odd part contains the terms with an odd number of $\gamma$ matrices. After taking the trace over spinor indices, this part becomes proportional to the Levi-Civita tensor $\epsilon^{\mu\nu\rho\sigma}$. On symmetry grounds, we may write this part as
\begin{equation}
S_{\rm odd}=\int d^4x\, d^4y\, F(x,y) A_\mu(x)\partial_\nu^y A_\rho(y)\epsilon^{\mu\nu\rho 3},\label{Sodd}
\end{equation}
where $F(x,y)$ is a nonlocal form factor depending on $x^3$, $y^3$ and $z^\alpha=x^\alpha-y^\alpha$, $\alpha=0,1,2$. After changing the integration variables, we write
\begin{equation}
S_{\rm odd}=\int d^3z^\alpha d^3y^\alpha dx^3 dy^3\, F(z^\alpha,x^3,y^3) A_\mu(z^\alpha+y^\alpha,x^3)\partial_\nu^y A_\rho(y^\alpha,y^3)\epsilon^{\mu\nu\rho 3},\label{Sodd2}
\end{equation}

When $\varphi\neq 0$, the theory has a mass gap. Therefore, the form factor $F(x,y)$ vanishes for large separations $|x-y|$. Also, since the action (\ref{Sodd}) is induced by the presence of the domain wall, the form factor has to vanish far away from the wall (for large $x^3$ or $y^3$). In the long wavelength limit, when the localization length of $F(x,y)$ is smaller than the characteristic scale of variation of $A_\mu$, the integral in (\ref{Sodd2}) factorizes as
\begin{equation}
S_{\rm odd}= \frac{ke^2}{4\pi} \int d^3y^\alpha A_\mu (y^\alpha,0)\partial_\nu A_\rho(y^\alpha,0) \epsilon^{\mu\nu\rho 3} \label{Sodd3}
\end{equation}
where
\begin{equation}
\frac{ke^2}{4\pi}=\int d^3z^\alpha dy^3 dx^3 F(z^\alpha,x^3,y^3)
\end{equation}
Thus, in the limit considered, the parity odd part of the effective action takes the form of the Chern-Simons action at the location of the domain wall.
$k$ is called the level of the Chern-Simons action.

The time component of the current corresponding to the action (\ref{Sodd}) reads 
\begin{equation}
J^0(x)=\frac 1e\, \frac \delta{\delta A_0(x)} S_{\rm odd}=
\frac 2e \int d^4y\, F(x,y) \partial_\nu^y A_\rho(y)\epsilon^{0\nu\rho 3}
\label{J0}
\end{equation}
By integrating this current\footnote{To the 2nd order in $A_\mu$, the parity even part of the effective action may be written as $S_{\rm even}=\int d^4xd^4y G(x,y)F_{\mu\nu}(x)F^{\mu\nu}(y)$ with some nonlocal kernel $G(x,y)$. The time component of the corresponding current is proportional to $F^{0\mu}$. This component of the field strength vanishes on the backgrounds that we consider here.} over spatial coordinates, one obtains the fermion number. In the long wave length approximation it may be written as
\begin{equation}
N=\int d^3x J^0(x)=\frac {ek}{2\pi} \int F_{12} d^2x \label{NJ}
\end{equation}
By comparing this formula with (\ref{Nfin}) we conclude
\begin{equation}
k=-\frac {\theta^+-\theta^-}{2\pi}\,.\label{k}
\end{equation}
Earlier this expression for the level of induced Chern-Simons action was conjectured in \cite{Mulligan:2013he} relying on analogies with lower dimensional models.

With the same formulas we can also evaluate the Chern-Simons terms induced on a boundary. Let us consider the scalar fields with a step-function profile,
\begin{eqnarray}
&&\varphi_{1,2}(x)=\varphi_{1,2}^-\quad \mbox{for}\quad x^3<0, \nonumber\\
&&\varphi_{1,2}(x)=\varphi_{1,2}^+\quad \mbox{for}\quad x^3>0.\label{step} 
\end{eqnarray}
(Later we shall take the limit $|\varphi^-|\to\infty$.) With a finite $|\varphi^-|$, let us consider the eigenmodes of $H$, $H\psi =E\psi$, without electromagnetic field, $A_\mu=0$, and vanishing at $x^3\to -\infty$. At $x^3<0$ such modes may be taken proportional to $e^{\kappa x^3+ik_ax^a}$, $x^a\in \{ x^1,x^2\}$ and $\kappa>0$. The parameters are restricted by the dispersion relation
\begin{equation}
E^2=-\kappa^2+k^2+|\varphi^-|^2 .\label{disp}
\end{equation}
Let us now take $|\varphi^-|\to\infty$ while keeping $E$ and $k_a$ finite. The dispersion relation (\ref{disp}) gives $\kappa \simeq |\varphi^-|$. The equation $H\psi=E\psi$ yields
\begin{equation}
\bigl( -\ii \alpha^3 +\beta e^{\ii \theta^- \gamma^5}\bigr)\psi =0.\label{chib1}
\end{equation}
This equation has to be satisfied for $x^3<0$ and, by continuity, it becomes a boundary condition at $x^3=0$ for the modes at the half space $x^3>0$. More commonly, it is written as a chiral bag boundary condition \cite{Rho:1983bh}
\begin{equation}
\bigl( 1 -\ii \gamma^3 e^{\ii \theta^- \gamma^5} \bigr) \psi\vert_{x^3=0}=0.\label{chib2}
\end{equation}
An interpretation of bag boundary conditions with $\theta^-=0$ through a singular limit of the scalar field $\varphi^-_1\to\infty$ was suggested in \cite{Chodos:1974je,Berry:1987qi}. The electromagnetic potential is smooth and thus does not influence the boundary condition (\ref{chib2}).

Hence, (\ref{k}) describes also the level of induced Chern-Simons action on a manifold with boundary, where $\theta^-$ defines the chiral phase in boundary conditions.

The restriction (\ref{step}) to constant values of $\varphi_{1,2}$ at positive $x^3$ is not essential and can be lifted. It is, however, useful to make a comparison to the computations \cite{Kurkov:2017cdz,Fialkovsky:2019rum} of Chern-Simons terms on the boundary (all done for non-chiral bag condition with $\theta^-=0$). The paper \cite{Kurkov:2017cdz} dealt with the massless case $|\varphi^+|=0$, where our formulas cannot give a unique answer (cf. the comment at the end of Sec.\ \ref{sec:fer}). The work \cite{Fialkovsky:2019rum} computed the polarization tensor when both $\varphi_1^+$ and $\varphi^+_2$ are non-zero constants. The results seem to be consistent but after an additional Pauli-Villars (PV) subtraction. Note, that the PV subtraction and the sharp boundary limit do not commute.

\section{Conclusions}\label{sec:con}
In this work we have computed the fermion number of a domain wall between two regions with  asymptotically constant $\varphi_1$ and $\varphi_2$. The fermion number appeared to be proportional to the difference of chiral phases in two asymptotic regions and to the magnetic flux passing through the wall. We used the method \cite{Alonso-Izquierdo:2019tms} based on a resummation of the heat kernel expansion. We kept only those terms in the heat kernel coefficients that contain a small number of derivatives. However, since the result is \emph{topological}, it is valid on any background having the same asymptotics. In the bulk of the manifold the fields may vary arbitrarily fast.

The same computation allowed us to evaluate the induced Chern-Simons action on interfaces. The results is consistent with the previous calculations \cite{Mulligan:2013he}\footnote{The paper \cite{Mulligan:2013he} used the PV regularization scheme and found that to make the induced Chern-Simons term finite one needs an extra (linear in masses) restriction on the PV regulators. This restriction is incompatible with the usual conditions appearing in the parity even part of polarization tensor in empty space. A similar condition was found in \cite{Fialkovsky:2019rum} in the presence of a boundary. In this latter paper all conditions on the PV regulators were made consistent by interchanging the roles of normal and axial regulator masses.} that were done for a fixed profile of $\varphi_1$ and a constant $\varphi_2$. By taking a singular limit we were able to compute the induced Chern-Simons action on a boundary with chiral bag boundary conditions.  In a particular case the result is consistent with the computations of \cite{Fialkovsky:2019rum}. We like to stress that the parity odd effective action (\ref{Sodd}) is not topological. There are derivative corrections to the long wavelength limit (\ref{Sodd3}) which cannot be computed from the fermion number (\ref{Nfin}).

\subsection*{Acknowledgments}
One of the authors (D.V.) is grateful to Max Kurkov for correspondence regarding induced Chern-Simons terms.
This work was supported in parts by the S\~ao Paulo Research Foundation (FAPESP), projects 2016/03319-6 and 2017/50294-1 (SPRINT), by the grants 303807/2016-4 and 428951/2018-0 of CNPq, by the RFBR project 18-02-00149-a and by the Tomsk State University Competitiveness Improvement Program. JMG also thanks the JCyL and the Spanish Governmemt-MINECO for partially supporting his research under the Grants BU 229P18, VA 057U16, SA 967G 19 and MTM 2014-57129-C2-1-P.

\end{document}